\documentclass[runningheads]{llncs}
\usepackage{graphicx}

\usepackage{amsmath,graphicx}
\usepackage{authblk}
\usepackage{amssymb}
\usepackage{multirow}
\usepackage[normalem]{ulem}
\usepackage[ruled,vlined]{algorithm2e}
\usepackage{soul}
\usepackage{subcaption}

\usepackage[colorlinks=true, linkcolor=black, urlcolor=blue]{hyperref}
\usepackage{xparse}
\usepackage{orcidlink}
\usepackage{todonotes}

\usepackage{cleveref}
\crefformat{footnote}{#2\footnotemark[#1]#3}
\begin{document}
\title{Beyond Pixels: Medical Image Quality Assessment with Implicit Neural Representations}
\author{Caner Özer\inst{1,2}
\and
Patryk Rygiel\inst{2}
\and
Bram de Wilde\inst{2}
\and İlkay Öksüz\inst{1}
\and Jelmer M. Wolterink\inst{2}
}
\authorrunning{Ozer et al.}
%
\institute{Computer Engineering Department, Istanbul Technical University, 34467 Maslak, Istanbul, Türkiye \email{\{ozerc\}[at]itu.edu.tr}
\and
Department of Applied Mathematics, Technical Medical Centre, University of Twente, Enschede, 7522 NB, The Netherlands}

\titlerunning{Beyond Pixels}
%

%
\maketitle              
\begin{abstract}
Artifacts pose a significant challenge in medical imaging, impacting diagnostic accuracy and downstream analysis. While image-based approaches for detecting artifacts can be effective, they often rely on preprocessing methods that can lead to information loss and high-memory-demand medical images, thereby limiting the scalability of classification models. In this work, we propose the use of implicit neural representations (INRs) for image quality assessment. INRs provide a compact and continuous representation of medical images, naturally handling variations in resolution and image size while reducing memory overhead. We develop deep weight space networks, graph neural networks, and relational attention transformers that operate on INRs to achieve image quality assessment. Our method is evaluated on the ACDC dataset with synthetically generated artifact patterns, demonstrating its effectiveness in assessing image quality while achieving similar performance with fewer parameters.

\end{abstract}

\section{Introduction}
Artifacts in cardiac magnetic resonance imaging (MRI) \cite{Ferreira2013} remain a persistent challenge, degrading diagnostic accuracy and complicating automated analysis. In cine MRI, artifacts typically arise from patient or respiratory motion \cite{Wang2022}, poor ECG triggering \cite{Lyu2021,Ozer2021}, or pacemakers \cite{Vuorinen2023}. Conventional image-based artifact detection methods \cite{Oksuz2019,Lei2022}, which are often heuristic-driven and parameter-heavy, struggle with scalability and consistency across different datasets.

Implicit neural representations (INRs) offer a promising alternative to commonly used voxel representations, providing compact, continuous signal representations that reduce memory demands while preserving essential details. INRs have shown success in tasks like periodic motion estimation \cite{Garzia2024}, cardiac segmentation \cite{Stolt-Anso2023}, and CT registration \cite{Wolterink2022}. Moreover, INRs have been widely used in MRI reconstruction, e.g. \cite{Chu2025}. This results in images that are represented in an INR. However, it's unclear how the quality of such an image can be directly assessed without having to resort to voxelization.

Clinically, distinguishing between moderate and severe artifacts is crucial, as these have the most significant impact on diagnostic reliability \cite{Loizillon2024}. Motivated by this, we develop a framework to detect mild and severe artifacts in cine cardiac MRI represented as INRs. We address the computational burden of INR optimization \cite{Papa2024} by introducing a parallel parameterization strategy that improves GPU efficiency. Our approach uniquely classifies artifact severity directly from INR parameters, combining parallel INR synthesis with weight-space classification, and represents, to our knowledge, the first such effort in medical image quality assessment.

\section{Methods}

\begin{figure}
    \centering
    \includegraphics[width=0.7\linewidth]{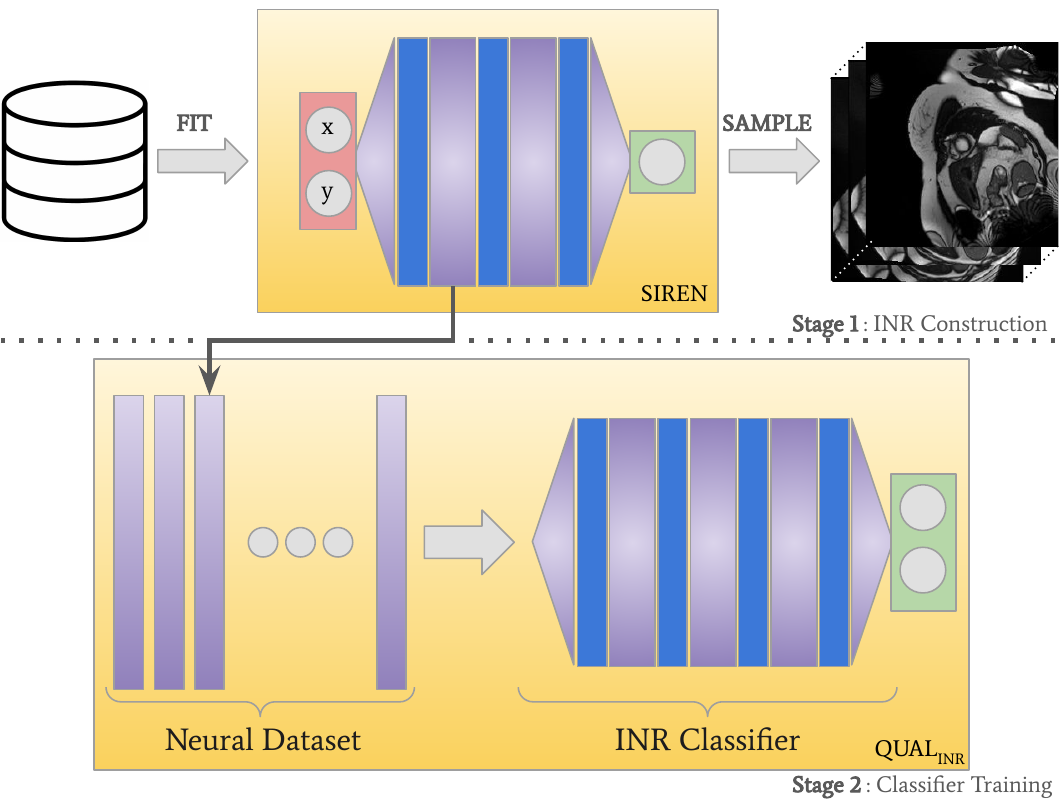}
    \caption{INR construction and weight space training for classifying good and mid/poor quality acquisitions. In Stage 1, 2D slices from cardiac MR volumes in the dataset are fitted into INRs, which can then be used to reconstruct the 2D image slices via sampling. In Stage 2, the parameters of these INRs are used for training and prediction without sampling the 2D image slices.
    }
    \label{fig:weightspaces}
\end{figure}

In this section, we present our methodology for quality assessment. As shown in Figure \ref{fig:weightspaces}, it consists of two stages. In Stage 1 (INR Construction), SIREN-based INRs are fitted to 2D slices of 3D+T cardiac MR images to encode them in weight space. This mimics the acquisition and direction reconstruction of MRI images to an INR. In Stage 2 (Classifier Training), these weights are used to train a classifier to predict image quality.

\subsection{Synthetic Mistriggering Artifact Generation Method}

We generate synthetic artifacts by modifying the k-space representation of cine sequences, swapping frequencies between adjacent cardiac phases. The proportion of swapped frequencies is controlled by a hyperparameter $P$, called the corruption ratio. Higher $P$ values lead to more pronounced mistriggering artifacts. We work with 2D + t image slices extracted from a specific depth of a 3D + t CMRI volume.

Let $N$ denote the number of frequencies in an image slice and $T$ the total number of adjacent slices. Given a 2D+time patient image, $I$, Algorithm \ref{algo:k_space} outlines the pseudocode for generating k-space-based mistriggering artifacts.

First, the algorithm computes the total number of frequencies to alter, $n'$, and randomly selects $n'$ frequencies from a uniform distribution over $[0, N-1]$. It also constructs a discrete reciprocal normal distribution over $[0, T-1]$ centered on the current time-slice $t$, to ensure source frames differ from $t$. Finally, the selected k-space lines are copied from these source time-frames to the target.

\begin{algorithm}
\SetAlgoLined
\KwResult{$I'$, Images with mistriggering artifact }
 $I$, $N$, $T$, $P$, $t$, $t'$, $n'$, $I'$\\
 \While{$t \leq T-1$}{
    $n'$ := $NP$ \\
    \textbf{select} random $n'$ lines from a Uniform distribution\\
    \textbf{form} a discrete reciprocal Normal distribution\\
    \textbf{select} random $t'$ lines from that distribution\\
    \textbf{apply} changes on $I'$ using $I$, $n'$ and $t'$\\
    $t := t + 1$
 }

 \caption{k-space based randomized mistriggering artifact generation procedure}
\label{algo:k_space}
\end{algorithm}

To generate images of varying quality, we set $P=0.05$ for mid-quality and $P=0.15$, for poor-quality images.

\subsection{Stage 1: INR Generation}

Given that we have good, mid, and poor-quality cardiac MRI acquisitions after applying our synthetic artifact generation method, we synthesize implicit neural representations (INRs) corresponding to the 2D slices present in these acquisitions. Formally, let us denote the set of all 2D images as signals $\{s_1, s_2, ..., s_N \}$. For each signal $s_i$, the coordinates $\textbf{x}_j \in \mathbb{R}^2$ and corresponding pixel intensity values $y_j\in \mathbb{R}$ are available, with a total of $M$ such coordinate-intensity pairs. The mapping $f_{\theta_i}: \mathbb{R}^2 \xrightarrow{} \mathbb{R}$ aims to model this relationship by optimizing the parameters $\theta_i$ for each signal $s_i$. 

We apply an $L$-layer MLP architecture with a hidden dimension of $D$ and sinusoidal activations of SIREN \cite{Sitzmann2019}, where the INR model is represented as: 

\begin{equation}
    f_{\theta_i}(\textbf{x}) = \textbf{W}_{l-1} (\phi_{l-2} \circ \phi_{l-3} \circ ... \circ \phi_0) + \textbf{b}_{l-1}, \quad \phi_l(\textbf{x}_i) = \sin(\omega_0(\textbf{W}_l \textbf{x}_i + \textbf{b}_l))
\end{equation}

where $\textbf{W}_{i}$ and $\textbf{b}_{i}$ are the learnable parameters of the $i$-th layer, and $\omega_0$ is a hyperparameter that controls the spatial frequency. Following \cite{Papa2024}, we apply a fixed initialization with a single seed across all $f_{\theta_i}$'s.

To optimize the parameters of $f_{\theta_i}$, we define a loss function that minimizes the squared error between the sampled and actual pixel intensity values for all available coordinate-intensity pairs:

\begin{equation}
    \mathcal{L} = \frac{1}{M} \sum_j \| y_j - f_{\theta_i}(\textbf{x}_j) \|^2 . 
\end{equation}

Since sequential optimization would take weeks with $2,000$ iterations per INR, we parallelize using JAX (v0.4.23) \cite{JAX} with vectorized mapping, fitting up to 400 INRs at a time on four NVIDIA PASCAL GPUs to fully utilize VRAM.

Overall, the SIREN-based network effectively models the quality of good-, mid-, and poor-quality images. We perform this fitting for all 2D slices, building a neural dataset from the INR parameters \footnote{\url{https://github.com/canerozer/fit-a-nef/}}.

\subsection{Stage 2: INR Classification}

This section details the INR classification after constructing the neural dataset. To preserve efficiency in terms of parameter utilization, we do not sample the image for classification purposes; instead, we work directly with the learned parameters of the INRs.

To evaluate different modeling strategies for INR classification, we assess four architectures: multi-layer perceptron (MLP), deep weight space networks (DWSNets) \cite{Navon2023}, graph neural networks (GNNs) \cite{Kofinas2024,Corso2020}, and relational attention transformers \cite{Kofinas2024,Diao2023}. Each of these architectures provides a unique perspective on learning representations from INR parameters. The MLP serves as a baseline approach, where the flattened INR parameters are treated as a high-dimensional input vector. DWSNets introduce equivariant layers that exploit the relational structure among INR weights and biases. GNNs are employed under the hypothesis that INR parameter structures exhibit relational properties, where connections between layers can be exploited to improve classification when represented as a graph. In this formulation, biases of INRs are encoded as node features, while weights act as edge features. Finally, relational transformers leverage the same structural property while aiming to model higher-order dependencies within INR weight distributions, capturing long-range interactions that may be overlooked by conventional architectures.  

\textbf{MLP:} The MLP consists of four hidden layers, each with 64 dimensions. The output of each layer is passed through a 1D batch normalization layer followed by ReLU activations. A final output layer completes the classification process.

\textbf{DWSNets:} The DWSNet implementation follows \cite{Navon2023}, incorporating four equivariant layers with 64 dimensions, along with a final invariant layer responsible for flattening, max-reduction, and prediction operations. \footnote{\url{https://github.com/canerozer/DWSNets}}

\textbf{GNN:} The GNN model is adopted from \cite{Kofinas2024}, utilizing four message-passing layers, each with a hidden size of 32. Additionally, 32-dimensional probing features are incorporated to augment the node representations. \footnote{\label{note1}\url{https://github.com/canerozer/NeuralGraphs}}

\textbf{Transformer:} The transformer model is based on \cite{Kofinas2024}, employing four attention layers with eight heads per layer, while other hyperparameters remain unchanged. Similar to the GNN, the transformer architecture also benefits from the probing mechanism, enhancing feature representation and learning. \cref{note1}

The classification tasks are divided into two separate evaluations: \textit{Good vs. Poor} and \textit{Good vs. Mid}. Both tasks involve predicting the quality category of an MRI slice based on the INR parameters learned during Stage 1. Given that these quality labels originate from controlled synthetic artifact generation, the classifier aims to identify INR-specific patterns that correspond to these predefined categories. We adopt binary cross-entropy loss as the optimization objective and use AdamW to optimize the model parameters. All of these models were implemented in PyTorch (v2.0.1). 

To prevent overfitting \cite{Shamsian2023} and improve generalization, we apply standard regularization techniques, including dropout ($0.2$, except for MLP), weight decay ($0.0005$), and data augmentation strategies such as rotation, noise injection, and scaling. Additionally, we employ early stopping based on validation accuracy, training the networks for a maximum of 10 epochs. The learning rate is set to $0.0005$ for MLP and DWSNet, $0.001$ for GNN and Transformer.

\section{Experimental Results}

In this section, we describe the dataset and present the experimental results of the classifiers. We evaluate the performance using classification accuracy for both the Good vs. Poor and Good vs. Mid tracks, repeating all experiments 10 times with different seed values. We will make the code available upon the acceptance of the paper. 

\textbf{Dataset:} We present our results using the ACDC Challenge dataset \cite{Bernard2018}, which comprises 150 3D+t short-axis Cardiac MR acquisitions. These acquisitions are evenly distributed across five cardiac conditions: normal, heart failure with infarction, dilated cardiomyopathy, hypertrophic cardiomyopathy, and abnormal right ventricle. To maintain a balanced representation, we allocate 90 acquisitions for training, 10 for validation, and 50 for testing. Synthetic artifacts are introduced following this distribution. During neural dataset construction, we apply center crops of $256 \times 256$ while ensuring that essential cardiac structures remain visible. After constructing the neural dataset, the sample distribution per quality label (good, mid, and poor) is as follows: $22,749$ for training, $2,602$ for validation, and $12,995$ for testing. Also, setting the number of layers, $L=3$, and their hidden dimension to $64$ allowing us to represent a signal with $14.8$ times fewer parameters considering a reduction from $65,536$ to $4,417$.

\subsection{Image Quality vs. INR Classification}

\begin{figure}
    \centering
    \includegraphics[width=0.6\linewidth]{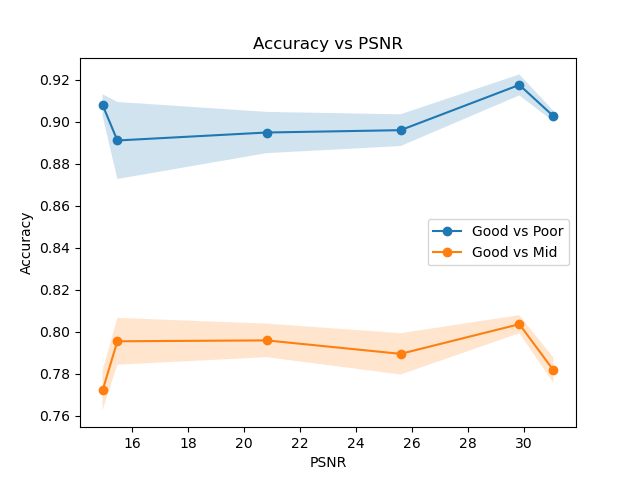}
    \caption{Relationship between reconstruction quality (PSNR) and INR classification performance for the MLP model. Each point shows the average classification accuracy at a PSNR threshold, with shaded standard deviation.}
    \label{fig:psnr_vs_acc}
\end{figure}

\begin{figure}
    \centering
    \includegraphics[width=\linewidth]{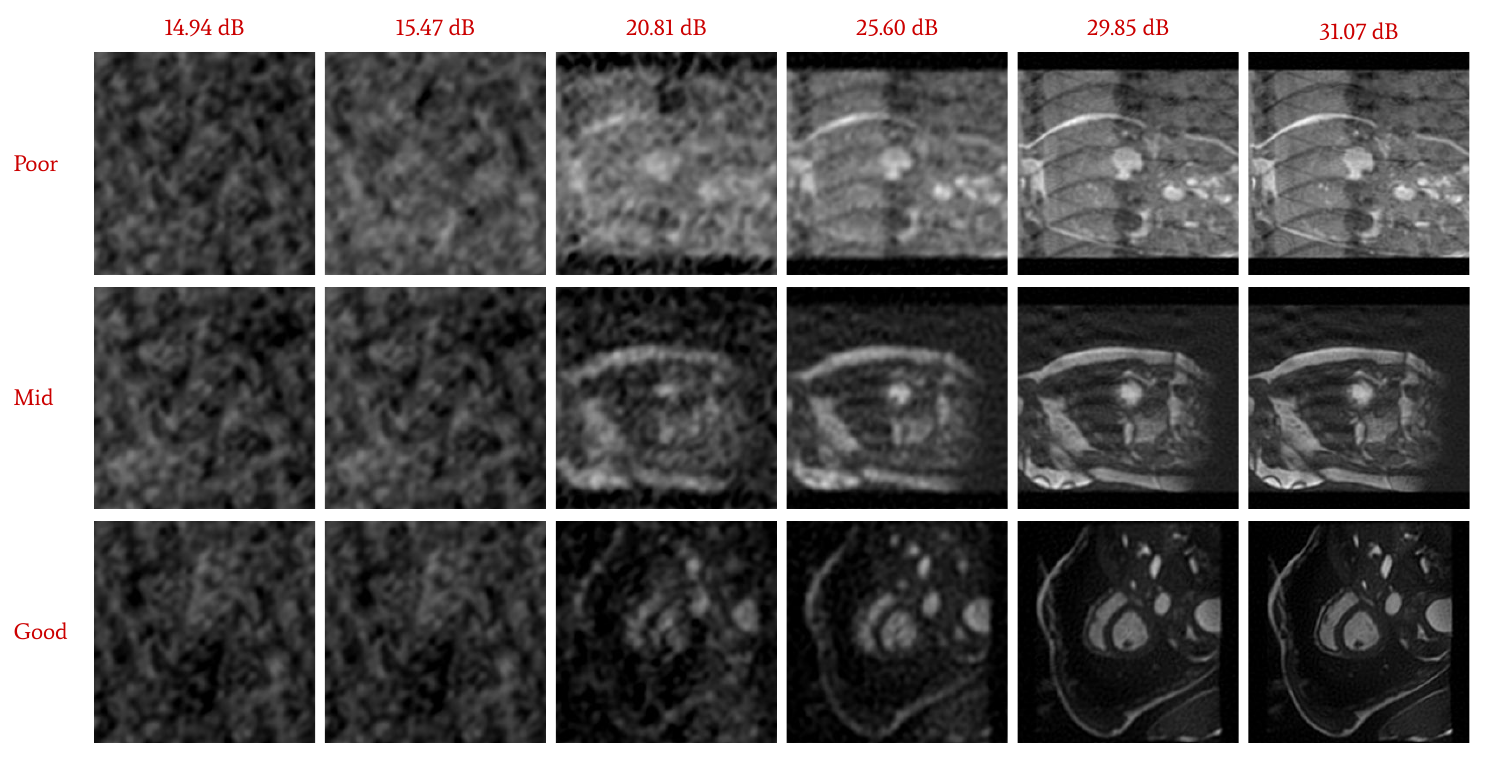}
    \caption{Signal reconstructions from various INRs trained on neural datasets with different average reconstruction PSNRs after early stopping. 
    }
    \label{fig:psnr_samples}
\end{figure}

Figure \ref{fig:psnr_vs_acc} presents the classification accuracy as a function of PSNR, where PSNR values are obtained through early stopping during INR optimization. This is done by defining multiple PSNR threshold values and halting INR parameter optimization once the corresponding threshold is reached. To see this effect of early stopping qualitatively, Figure \ref{fig:psnr_samples} underscores this by illustrating how the reconstructed signals qualitatively vary across different average PSNR levels after early stopping, providing visual examples that complement the quantitative accuracy trends in Figure \ref{fig:psnr_vs_acc}. 


We observe that the accuracy for Good vs. Poor remains relatively stable across PSNR levels but shows an upward trend at higher PSNR values, indicating that improved INR reconstructions enhance classification reliability. In contrast, Good vs. Mid classification exhibits slightly more fluctuations, with accuracy decreasing at intermediate PSNR values before improving at higher PSNR levels. 

The best classification performance is observed at a PSNR of $29.85$ dB, where accuracy reaches $0.918$ for Good vs. Poor and $0.804$ for Good vs. Mid. However, when training continues for $2,000$ iterations, resulting in an average PSNR of $31.07$ dB, accuracy slightly decreases to $0.904$ for Good vs. Poor and $0.787$ for Good vs. Mid. This suggests that while improving INR reconstruction quality generally enhances classification, excessive training may introduce minor degradation in performance. Overall, the trend indicates that higher PSNR, achieved through better INR optimization, contributes to improved classification accuracy, particularly for the Good vs. Poor task.

An interesting observation from this plot is that an average reconstruction quality of $14.97$ dB of PSNR is achieved after just a single iteration, while still obtaining an average accuracy of $0.9081$ for Good vs. Poor and $0.7721$ for Good vs. Mid. This suggests that even the early-stage INR parameters provide predictive information to enable classification to some extent. We focus on converged INR parameters, as they are likely to be more informative for applications beyond classification.

\subsection{Image Quality Assessment Performance}
In Table \ref{tab:cls_res}, we present the results of the test splits. The first row corresponds to image-based classification using a pixel-based MLP classifier operating on flattened pixel intensities, with over 12 million parameters due to the input resolution of $256 \times 256$. Despite its high parameter count, the pixel-based MLP underperforms compared to INR-based methods, achieving only 0.727 and 0.529 accuracy for the Good vs. Poor and Good vs. Mid tasks, respectively. This performance gap suggests that flattened pixel intensities alone may not sufficiently capture essential spatial patterns indicative of image quality, highlighting the need for more effective representations.

Among the INR-based methods, the Transformer (Probe=$32$) model achieves the highest performance, with classification accuracies of $0.944$ and $0.833$ on the Good vs. Poor and Good vs. Mid tasks, respectively, while using significantly fewer parameters than the image-based MLP model. Notably, even the INR-based MLP model achieves accuracies of $0.904$ and $0.787$ for the same tasks. These results highlight both the effectiveness of lightweight architectures operating in INR weight space and the benefits of probe-based mechanisms. Furthermore, the superior performance of the Transformer (Probe=$32$) model is statistically significant compared to other models, with a p-value $< 0.01$.


\begin{table}[]
\centering
\begin{tabular}{|l|c|c|c|}
\hline
\hline
{\color[HTML]{000000} \textbf{Model Name}}      & {\color[HTML]{000000} \textbf{\begin{tabular}[c]{@{}c@{}}Number of\\ Parameters\end{tabular}}} & {\color[HTML]{000000} \textbf{Good vs. Poor}}  & {\color[HTML]{000000} \textbf{Good vs. Mid}}   \\ \hline \hline
Image-Based Model (MLP)                              & $12.6$M                                   & $0.727$                                         & $0.529$                                         \\ \hline \hline
{\color[HTML]{000000} MLP}                      & {\color[HTML]{000000} $300$K}             & {\color[HTML]{000000} $0.904$}                  & {\color[HTML]{000000} $0.787$}                  \\ \hline
{\color[HTML]{000000} DWSNet}                   & {\color[HTML]{000000} $2.16$M}            & {\color[HTML]{000000} $0.901$}                  & {\color[HTML]{000000} $0.770$}                  \\ \hline
{\color[HTML]{000000} GNN}                      & {\color[HTML]{000000} $96$K}              & {\color[HTML]{000000} $0.929$}                  & {\color[HTML]{000000} $0.808$}                  \\ \hline
{\color[HTML]{000000} Transformer}              & {\color[HTML]{000000} $376$K}             & {\color[HTML]{000000} $0.933$}                  & {\color[HTML]{000000} $0.816$}                  \\ \hline
{\color[HTML]{000000} GNN (Probe=$32$)}           & {\color[HTML]{000000} $96$K}              & {\color[HTML]{000000} $0.938$}                  & {\color[HTML]{000000} $0.817$}                  \\ \hline
{\color[HTML]{000000} Transformer (Probe=$32$)}   & {\color[HTML]{000000} $384$K}                                                                    & {\color[HTML]{000000} {\textbf{0.944}}}  & {\color[HTML]{000000} { \textbf{0.833}}} \\ \hline \hline
\end{tabular}\caption{Performance comparison of various image-based and INR-based models on the "Good vs. Poor" and "Good vs. Mid" quality classification.}
\label{tab:cls_res}
\end{table}

\section{Conclusion \& Discussion}
This study introduced an INR-based framework for medical image quality assessment, leveraging neural fields to efficiently classify mistriggering artifacts in cardiac MRI. By directly using INR parameters, our method reduces memory and computational overhead compared to traditional image-based approaches, while maintaining high accuracy. Parallelized INR optimization further supports scalability across large datasets.

Our experiments in the ACDC dataset demonstrated that INR-based classification achieves improved accuracy compared to the image-based MLP model, with the Transformer (Probe=32) achieving 94.4\% (Good vs. Poor) and 83.3\% (Good vs. Mid) accuracy. We also notice that classification performance improved with INR reconstruction quality, peaking at 29.85 dB PSNR, though even early-stage INR optimization retained strong predictive power. These findings highlight the potential of INR parameters as meaningful descriptors of image quality, supporting the feasibility of neural fields for artifact detection.

Limitations include the computational cost of INR optimization, even in parallel, and the restricted variability of synthetic artifacts, which may not fully reflect real-world cases. Broader validation on datasets like CMRxMotion \cite{Wang2022} and LDCTIQAC \cite{Lee2025} is left for future work.

Overall, our approach demonstrates a promising, lightweight alternative to large CNN-based models for scalable artifact detection. Future directions include improving INR optimization speed for real-time use, broader dataset validation, and exploring INR editing \cite{Zhou2023} for artifact correction to further support clinical decision-making.

\subsubsection{\ackname} This work has been benefitted from the 2232 International Fellowship for Outstanding Researchers Program of TUBITAK (Project No: 118C353). However, the entire responsibility of the thesis belongs to the owner. The financial support received from TUBITAK does not mean that the content of the thesis is approved in a scientific sense by TUBITAK. This work was supported by Scientific Research Projects Department of Istanbul Technical University. Project Numbers: 44250 and 47296.

\bibliographystyle{splncs04}
\bibliography{miccai2025}

\end{document}